\documentstyle[aps,multicol,pra,psfig]{revtex}
\begin{document}

\title{Physical Interpretation of
Laser-Induced Suppression of Quantum Tunneling}

\author{Emmanuel Paspalakis}

\address{Optics Section, Blackett Laboratory, 
Imperial College, London SW7 2BZ, United Kingdom}

\date{\today}
\maketitle
\begin{abstract}
We revisit the problem of laser-induced 
suppression of quantum dynamical tunneling in 
a model system studied
by Kilin {\it et al.} [Phys.\ Rev.\ Lett.\ {76}  (1996) 3297].
This quantum system consists of a 
ground state symmetric double-well potential which is coupled
by a strong laser field to an excited state asymmetric 
double-well potential. 
By analyzing the assumptions used in their analysis
we show that the  suppression of quantum dynamical  tunneling
can be explained with the use of dark and bright states
of the system. We also generalize the system
and the conditions for suppression
of quantum tunneling and show that, in certain cases, suppression
can occur regardless the characteristics of the excited potential surface.
\end{abstract}


\begin{multicols}{2}

One of the most interesting features of a $\Lambda$-type quantum system
(a system with two lower states and one upper state) 
driven by two laser fields
is the appearance of a dark eigenstate of the system under conditions
of two-photon resonance \cite{Shorebook,Arimondo96a}. The dark state is a 
linear combination of only the two lower states of the system,
and not the excited state. Dark states
also exist in systems where interfering dissipative processes are present
\cite{Knight90a}.
Numerous interesting phenomena
are associated with dark states, such as coherent population trapping
\cite{Trapping} and transfer \cite{Kuklinski89a,STIRAPrev},
electromagnetically induced transparency \cite{Eberly95a,Harris97a} and 
propagation of matched pulses \cite{Harris94a}, propagation of
soliton-like pulses in multi-level media (like ``simultons'' \cite{Stroud81a}
and ``adiabatons'' \cite{Grobe94a,Fleischhauer96a}), intrinsic transparency \cite{Paspalakisprl99a}, lasing without inversion \cite{LWI}, creation of 
radiation fields with matched photon statistics \cite{Agarwal93a,Fleischhauer94a},
quenching of spontaneous emission \cite{Agarwalbook,Zhu96a,Paspalakisprl98a}
or resonance fluorescence \cite{Hegerfeldt92a,Zhou96a}, velocity selective laser cooling \cite{Aspect88a} and others.

In a recent article, Kilin {\it et al.} 
\cite{Kilin96a} showed that it is possible
to manipulate coherently the process of quantum dynamical tunneling 
\cite{Hanggi98a}
in a model system consisting of a ground symmetric double-well
potential $V_{g}(x)$ and an excited
asymmetric double-well potential $V_{e}(x)$
(see Fig.\ 1).
It is well known (see, for example, 
ref. \cite{CohenTannoudjibook}) that if the 
wavepacket is initially 
localized in one of the two ground state potential wells, then 
it will tunnel through the barrier to the other well and 
will ultimately oscillate between the two potential wells.
If now the ground state well is coupled to the excited state well
by a strong laser field, this  will lead to the localization of the
wavepacket in one of the wells \cite{Kilin96a}, leading to suppression
of quantum dynamical  tunneling.
In this article, we show that the phenomenon of suppression of 
quantum tunneling which was discussed by Kilin {\it et al.} \cite{Kilin96a}
has its origin in the properties of the
$\Lambda$-type system. This is shown by analyzing the assumptions 
used in the above article and transforming the problem
to a basis of dark and bright states of the system \cite{Shorebook}. 
We also generalize the system and show that one 
assumption is required for the suppression
of quantum tunneling, if the
properties of the dark state is properly exploited.

We begin with an analysis of the process of quantum dynamical tunneling
in our model system
\cite{Hanggi98a,CohenTannoudjibook}. We consider only the ground state
symmetric double-well potential shown in Fig.\ 1 and recall that 
the wave functions of states
$|1\rangle$ and $|2\rangle$ can be written as
symmetric and antisymmetric superpositions of the localized
wave functions in each of the potential wells, 
\begin{eqnarray}
\psi_{1}(x) = \frac{1}{\sqrt{2}}\left[\phi(x) + \phi(-x) \right]
\, , \\
\psi_{2}(x) = \frac{1}{\sqrt{2}}\left[\phi(x) - \phi(-x) \right]
\, .
\end{eqnarray}
Here, $\psi_{1}(x) = \langle x|1\rangle$, $\psi_{2}(x) = \langle x|2\rangle$
and $\phi(x)$ $\left(\phi(-x)\right)$ is the wave function in the right-
(left-) side of the potential well. At $t=0$ we suppose that
the wave function of the system is written as a superposition
of the two states $|1\rangle$ and $|2\rangle$,
$|\psi(t=0)\rangle = c_{1}|1\rangle + c_{2}|2\rangle$, leading to
\begin{equation}
\psi(x,t=0) = \frac{c_{1}+c_{2}}{\sqrt{2}}\phi(x) +
\frac{c_{1}-c_{2}}{\sqrt{2}}\phi(-x) \, ,
\end{equation}
with $\psi(x, t)= \langle x|\psi(t) \rangle$.
The evolution of the system at any time $t$ is given by 
(we use units such that $\hbar=1$) 
\begin{eqnarray}
\psi(x, t) &=& c_{1} \psi_{1}(x) e^{-i \omega_{1} t} + 
 c_{2} \psi_{2}(x) e^{-i \omega_{2} t} \, \nonumber \\
&=& \frac{e^{-i(\omega_{1}+\omega_{2})t/2}}{\sqrt{2}}
\bigg[ \left(c_{1}e^{i\delta t/2} + c_{2}e^{-i\delta t/2} \right) \phi(x)
\nonumber \\
&+& \left(c_{1}e^{i\delta t/2} - c_{2}e^{-i\delta t/2} \right) \phi(-x)
\bigg] \, ,
\end{eqnarray}
where $\omega_{i}$, $(i=1,2)$ is the energy of state
$|i\rangle$ and $\delta = \omega_{2} - \omega_{1}$ 
is the separation energy of states $|2\rangle$ and $|1\rangle$.
This energy separation depends on the width of the potential
barrier between the wells.
Therefore, if the wavepacket is initially localized
on the left $[\psi(x, t=0) = \phi(-x)]$, $(c_{1} = -c_{2} = 1/\sqrt{2})$
then
\begin{eqnarray}
\psi(x, t) &=& e^{-i(\omega_{1}+\omega_{2})t/2}
\bigg[ i \sin{(\delta t/2)} \phi(x) \nonumber \\
&+& 
\cos{(\delta t/2)} \phi(-x)
\bigg] \, ,
\end{eqnarray}
so that
\begin{equation}
P_{L}(t) = \cos^{2}{(\delta t/2)} \quad , \quad P_{R}(t) = \sin^{2}{(\delta t/2)} \, ,
\end{equation}
with $P_{L}(t)$ $\left(P_{R}(t)\right)$ being the probability for the wavepacket to be
localized on the left- (right-) side of the well.
If now the wavepacket is initially localized on the right
$[\psi(x, t=0) = \phi(x)]$, $(c_{1} = c_{2} = 1/\sqrt{2})$ then
\begin{eqnarray}
\psi(x, t) &=& e^{-i(\omega_{1}+\omega_{2})t/2}
\bigg[  
\cos{(\delta t/2)} \phi(x) \nonumber \\
&+& i \sin{(\delta t/2)} \phi(-x)
\bigg] \, ,
\end{eqnarray}
and
\begin{equation}
P_{L}(t) = \sin^{2}{(\delta t/2)} \quad , \quad P_{R}(t) = \cos^{2}{(\delta t/2)} \, .
\end{equation}
So, in both of the above cases the wavepacket will oscillate between the two 
potential wells and will be localized at each of the wells 
only at certain times $t = n \pi/\delta$, with $n$ being an integer. 

We now suppose that a coherent, step pulse laser field couples
states $|1\rangle$ and $|2\rangle$ with a state $|3\rangle$ belonging
to another potential well surface (see Fig.\ 1). 
The Hamiltonian of this system, in the the rotating wave approximation,
is given by 
\begin{eqnarray}
H &=& \sum^{3}_{i=1} \omega_{i} |i\rangle \langle i| + \bigg[\Omega_{1} e^{i \omega t} |1\rangle \langle 3|
\nonumber \\
&+& \Omega_{2} e^{i \omega t} |2\rangle \langle 3| + \mbox{H.c.} \bigg]
- i \frac{\gamma}{2} |3\rangle \langle 3| \, .
\end{eqnarray}
Here, $\Omega_{i} = - \mu E \int dx \psi^{*}_{i}(x) \psi_{3}(x)$,
$(i=1,2)$ is the Rabi frequency 
of the 
$|i \rangle \rightarrow |3 \rangle$
transition. The Rabi frequency has been
obtained using the adiabatic approximation and is
assumed to be real.
Also, $\mu$ is the electric dipole matrix element, 
$E$ is the electric field amplitude and $\omega$ is the 
angular frequency of the laser field. Finally, $\gamma$ denotes the
decay of the excited state, which is assumed to occur outside
of the system and has been added phenomenologically to the
Hamiltonian. The wave function of the system is 
expanded in terms of the ``bare'' state vectors as
$|\psi(t)\rangle = \sum^{3}_{i=1} a_{i}(t) |i\rangle$. We 
substitute the above Hamiltonian and wave function into
the time-dependent Schr\"{o}dinger equation and
after a transformation we obtain 
\begin{eqnarray}
i \dot{{\bf b}}(t) = {\bf H} {\bf b}(t)   \, ,
\end{eqnarray}
where ${\bf b}(t) = ({b}_{1}(t),
{b}_{2}(t), 
{b}_{3}(t))^{T}$ and
\begin{eqnarray}
{\bf H} = \left(
\begin{array}{ccc}
 \delta_{1} &  0   & \Omega_{1}\\
0  &  \delta_{2} &  \Omega_{2}  \\
\Omega_{1} & \Omega_{2}  &  - \frac{i}{2}\gamma \\
\end{array}
\right) \; \, ,
\end{eqnarray}
with $\delta_{i} = \omega - \omega_{3} + \omega_{i}$,
$(i=1,2)$ being the detuning of the 
$|i \rangle \leftrightarrow |3 \rangle$
transition
and $a_{1}(t) = b_{1}(t) e^{-i (\omega_{3} - \omega)t}$,
$a_{2}(t) = b_{2}(t) e^{-i (\omega_{3} - \omega)t}$,
$a_{3}(t) = b_{3}(t) e^{-i \omega_{3} t}$.
We also note that $\delta_{2} = \delta_{1} + \delta$.
The Hamiltonian ${\bf H}$ is the same as that used
by Kilin {\it et al.} 
\cite{Kilin96a} if $\gamma = 0$ (see Eq.\ (10) of ref.\ \cite{Kilin96a}).
It is also the Hamiltonian of a $\Lambda$-type atomic system driven
by a single laser field \cite{Shorebook,Arimondo96a,Trapping}. 

We now define the ``dark'' $|-\rangle$
and ``bright'' $|+\rangle$ states as
\begin{eqnarray}
|+\rangle = \frac{1}{\Omega} \left(\Omega_{1}|1\rangle + \Omega_{2}|2\rangle \right) \, , \\
|-\rangle = \frac{1}{\Omega} \left(\Omega_{2}|1\rangle - \Omega_{1}|2\rangle \right) \, ,
\end{eqnarray}
with $\Omega = \sqrt{\Omega^{2}_{1} + \Omega^{2}_{2}}$. In this basis
the equations for the probability amplitudes are written as
\begin{eqnarray}
i \dot{b}_{+}(t) &=& \frac{\delta_{1}\Omega^{2}_{1} +
\delta_{2}\Omega^{2}_{2}}{\Omega^{2}} b_{+}(t) +
\frac{(\delta_{1} - \delta_{2})\Omega_{1}\Omega_{2}}{\Omega^{2}} b_{-}(t)
\nonumber \\
&+&\Omega b_{3}(t) \, , \\
i \dot{b}_{-}(t) &=&  
\frac{(\delta_{1} - \delta_{2})\Omega_{1}\Omega_{2}}{\Omega^{2}} b_{+}(t)
+ \frac{\delta_{1}\Omega^{2}_{2} +
\delta_{2}\Omega^{2}_{1}}{\Omega^{2}} b_{-}(t) \, , \\
i \dot{b}_{3}(t) &=& -i \frac{\gamma}{2} b_{3}(t) + \Omega b_{+}(t) \, .
\end{eqnarray}
with
\begin{eqnarray}
b_{+}(t) = \frac{\Omega_{1}}{\Omega} b_{1}(t) + \frac{\Omega_{2}}{\Omega} b_{2}(t) \, , \\
b_{-}(t) = \frac{\Omega_{2}}{\Omega} b_{1}(t) - \frac{\Omega_{1}}{\Omega} b_{2}(t) \, .
\end{eqnarray}
Then, the wavefunction of the system can be written as
\begin{eqnarray}
|\psi(t)\rangle &=& \left[b_{+}(t)|+\rangle +b_{-}(t)|-\rangle\right]e^{-i (\omega_{3} - \omega)t} \nonumber \\
&+& b_{3}(t)e^{-i \omega_{3} t}|3\rangle \, . 
\end{eqnarray}
We immediately note that if $\delta_{2} \approx \delta_{1}$
state $|+\rangle$ couples
only to state $|3\rangle$ and state $|-\rangle$ does not couple
to either states $|+\rangle$ or $|3\rangle$.
Therefore, if $\gamma \neq 0$ states $|+\rangle$ and $|3\rangle$
decay but the dark state $|-\rangle$ does not. 
Condition $\delta_{2} \approx \delta_{1}$ is the well-known
dark state condition of the $\Lambda$-type
system \cite{Shorebook,Arimondo96a,Trapping}. 
Only when this condition
is satisfied does the Hamiltonian ${\bf H}$ have a real, stable 
eigenvalue and coherent population trapping is possible.

Kilin {\it et al.} \cite{Kilin96a} assumed that 
$\delta_{2} \approx \delta_{1}$,
or in other words that $\delta_{1} + \delta \approx \delta_{1}$.
(Note that after Eq.\ (10) in ref.\ \cite{Kilin96a}, $\delta$ does 
not appear in the equations.) In addition, they assumed that
the excited state belongs to an asymmetric quantum-well potential
so that its wave function is localized on the left
such that
\begin{eqnarray}
\Omega_{1} &=& - \mu E \int dx \psi^{*}_{1}(x) \psi_{3}(x)
\nonumber \\ &\approx& - \frac{\mu E}{\sqrt{2}} \int dx \phi^{*}(-x) \psi_{3}(x)  \, , \\
\Omega_{2} &=& -\mu E \int dx \psi^{*}_{2}(x) \psi_{3}(x)
\nonumber \\ &\approx&  \frac{\mu E}{\sqrt{2}} \int dx \phi^{*}(-x) \psi_{3}(x)
= -\Omega_{1} \, .
\end{eqnarray}
Finally, they considered only the case of a metastable 
excited state, i.e. that $\gamma = 0$.
With these assumptions $|+\rangle = (|1\rangle - |2\rangle)/\sqrt{2}$,
$\psi_{+}(x) = \langle x|+\rangle = \phi(-x)$
and $|-\rangle = (|1\rangle + |2\rangle)/\sqrt{2}$,
$\psi_{-}(x) = \langle x|-\rangle = \phi(x)$.
Hence, if the wavepacket is initially localized on the left 
it will remain 
localized on the left and will simply
oscillate between states $|+\rangle$ and $|3\rangle$.
If now the wavepacket is initially
localized on the right, it will remain on the right as
state $|-\rangle$ is uncoupled from states $|+\rangle$
and $|3\rangle$. Therefore, suppression 
of quantum dynamical tunneling occurs.

How important are the above approximations? 
If the upper state
is not metastable $(\gamma \neq 0)$ and the system is initially
localized on the left it will simply leave the system as the upper
state decays. However, if the system is initially
localized on the right it will remain there, because the dark
state $|-\rangle$ does not couple to the decaying state 
$|3\rangle$. If now the system starts from an arbitrary superposition
of $|1\rangle$ and $|2\rangle$, only the part of the superposition
related to state $|-\rangle$ will be localized on the right, i.e.
\begin{eqnarray}
P_{L}(t \rightarrow \infty) &=& 0 \, , \\ 
P_{R}(t \rightarrow \infty) &=& P_{R}(t) = \frac{1}{2}\left[1+2\mbox{Re}(c_{1}c^{*}_{2})\right] \, .
\end{eqnarray}
So, localization can occur even if the upper state decays out of the 
system.

When the excited state does not belong to an
asymmetric quantum well leading to $|\Omega_{1}| \neq |\Omega_{2}|$ and 
the system is initially prepared in the dark state
$|-\rangle$ then  a part of the wavepacket will remain localized
on the left and another part will remain localized on the right
of the lower potential well.
In addition, no decay will occur as in this case state
$|-\rangle$ does not couple to states $|+\rangle$ and $|3\rangle$. 
The localization probabilities
for the system initially in the dark state are given by
\begin{eqnarray}
P_{L}(t) = \frac{(\Omega_{1}+\Omega_{2})^{2}}{2\Omega^{2}} \, , \\
P_{R}(t) = \frac{(\Omega_{1}-\Omega_{2})^{2}}{2\Omega^{2}} \, .
\end{eqnarray}
For any other initial condition, the overlap of the initial
state with the dark state $|-\rangle$ will determine the degree
of localization. In this case decay will occur
out of the system and the localization probabilities will be given
by
\begin{eqnarray}
P_{L}(t \rightarrow \infty) = \left|\Omega_{2}c_{1}-
\Omega_{1}c_{2}\right|^{2}\frac{(\Omega_{1}+\Omega_{2})^{2}}{2\Omega^{4}} \, , \\
P_{R}(t \rightarrow \infty) = \left|\Omega_{2}c_{1}-
\Omega_{1}c_{2}\right|^{2}\frac{(\Omega_{1}-\Omega_{2})^{2}}{2\Omega^{4}} \, .
\end{eqnarray}
For example, if the wavepacket is initially localized on the left then
after a transient period where damped tunneling oscillations will occur
the system will be localized on both left- and right-side well potential
with probabilities 
\begin{eqnarray}
P_{L}(t \rightarrow \infty) =
\frac{(\Omega_{1}+\Omega_{2})^{4}}{4\Omega^{4}} \, , \\
P_{R}(t \rightarrow \infty) = \frac{(\Omega_{1}^{2}-\Omega^{2}_{2})^{2}}{4\Omega^{4}} \, .
\end{eqnarray}
Therefore, regardless of the shape of the upper potential well, 
localization of the wavepacket is possible.
Even for the case that the upper state is not bound but is a 
continuum, suppression of quantum tunneling could 
occur. In that case the problem reduces to that of laser-induced
continuum structure \cite{Knight90a}
and dark states that lead to 
wavepacket localization can occur in this case, too.

Obviously only the condition $\delta_{2} \approx \delta_{1}$ is crucial here because only under this assumption a
dark state is formed.
This condition can be fulfilled, for example
in molecular systems where 
 $\delta \approx 10^{-5} - 10^{-4}$ eV and $\delta_{1}$ or
$\delta_{2}$ could be a few orders of magnitude larger
for optical transitions. 

In conclusion, we have re-examined the prototype system of
Kilin {\it et al.} \cite{Kilin96a} for laser-induced
suppression of quantum tunneling. Using an analysis in dark
and bright states we have shown that there is only one
important condition for suppression of tunneling and localization
of the wavepacket. This condition leads to a dark state
in the system. Hence, the phenomenon of laser-induced suppression
of quantum dynamical tunneling should also be added
to the list of the
phenomena  
that occur via the
creation of dark states.
Before closing we note that quantum tunneling
oscillations, similar to that studied in this article,
have been predicted in Bose-Einstein 
condensates of atomic gases using double-well trapping
potentials \cite{BEC}.
It may be therefore possible to use the idea presented
here to suppress these oscillations.

E.P. would like to thank Niels J. Kylstra and  Peter L. Knight 
for helpful
discussions and useful comments on the manuscript.
This work has been supported by the United Kingdom Engineering and Physical
Sciences Research Council (EPSRC).

\end{multicols}

\pagebreak

\begin{figure}
\begin{center} 
\leavevmode 
\centerline{\hbox{
\psfig{figure=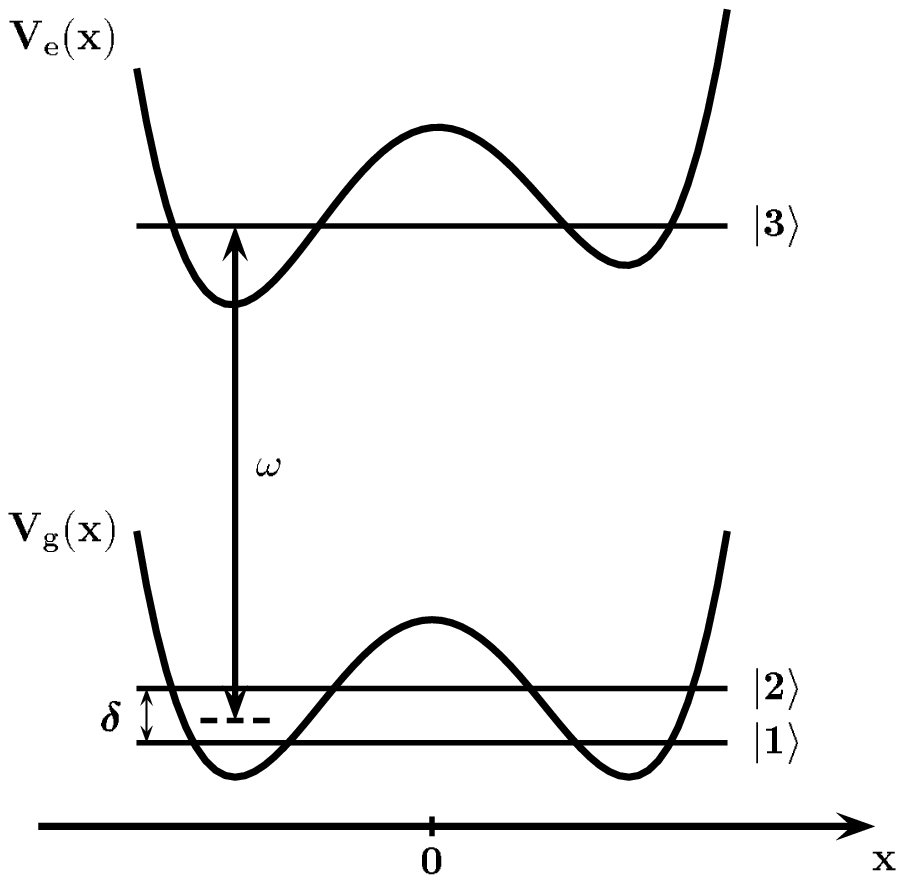,height=10.cm}}}
\end{center}
\caption
{  
The quantum system considered here consists of a ground symmetric double-well
potential $V_{g}(x)$ and an excited double-well potential $V_{e}(x)$. 
The excited potential is shown to be asymmetric;
however, as stated in the text, our results can be applied for more general
potentials too. States $|1\rangle$ and $|2\rangle$ are respectively the ground
and  first excited state of the lower potential and state $|3\rangle$ is 
a state of the upper potential. These states are coupled by a laser
field having angular frequency $\omega$.
}
\label{2afig1}
\end{figure}

\end{document}